\documentclass[a4paper]{jpconf}
\usepackage{url}
\usepackage{graphicx}
\begin{document}
\title{Heavy quark production at an Electron-Ion Collider}
\author{E~Chudakov$^1$, D~Higinbotham$^1$, Ch~Hyde$^2$, S~Furletov$^1$,
Yu~Furletova$^1$, D~Nguyen$^{1,3}$, M~Stratmann$^4$, M~Strikman$^5$,
C~Weiss$^1$}
\address{$^1$ Jefferson Lab, Newport News, VA 23606, USA \\
$^2$ Old Dominion University, Norfolk, VA 23529, USA \\
$^3$ University of Virginia, Charlottesville, VA 22904, USA \\
$^4$ T\"ubingen University, 72076 T\"ubingen, Germany \\
$^5$ Pennsylvania State University, University Park, PA 16802, USA}
\ead{weiss@jlab.org}
\begin{abstract}
An Electron-Ion Collider (EIC) with center-of-mass energies $\sqrt{s_{eN}} \sim$ 20--100 GeV 
and luminosity $L \sim 10^{34}$ cm$^{-2}$ s$^{-1}$ would offer new opportunities to 
study heavy quark production in high-energy electron or photon scattering on protons
and nuclei. We report about an R\&D project exploring the feasibility of direct 
measurements of nuclear gluon densities at $x > \sim 0.1$ (gluonic EMC effect, 
antishadowing) using open charm production at EIC. We describe the charm production rates 
and angle-momentum distributions at large $x$ and discuss methods of charm reconstruction 
using next-generation detector capabilities ($\pi/K$ identification, vertex reconstruction).
The results could be used also for other physics applications of heavy quark production 
at EIC (fragmentation functions, jets, heavy quark propagation in nuclei).
\end{abstract}
\section{Electron-Ion Collider}
\begin{figure}
\parbox[c]{0.55\textwidth}{
\includegraphics[width=0.55\textwidth]{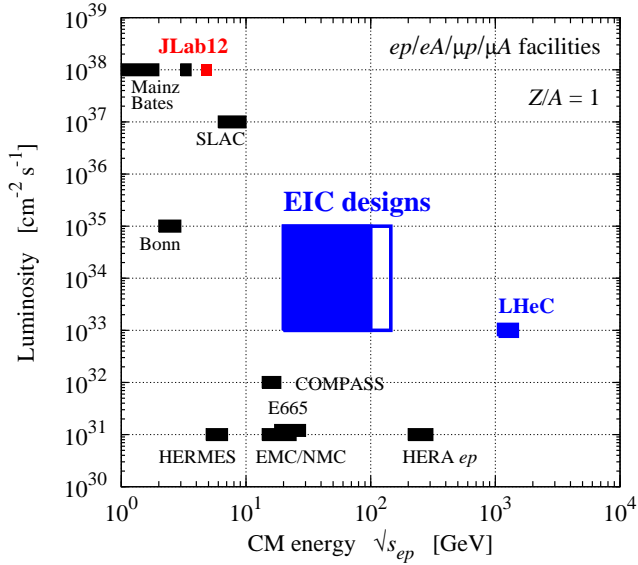}}
\hspace{0.04\textwidth} 
\parbox[c]{0.33\textwidth}{
\caption[]{CM energies and luminosities of experimental facilities for 
electromagnetic scattering at multi-GeV energies ($ep/eA/\mu p/\mu A$;
past, present, and future). 
The coverage of the EIC designs is indicated by the blue box
($ep, Z/A = 1$).
\label{fig:elumi}}}
\end{figure}

An Electron-Ion Collider (EIC) is being developed as a next-generation facility for 
nuclear physics and has been recommended for future construction in the 2015 
U.S.~Department of Energy's Long-Range Plan \cite{2015NSAC}. The present EIC designs
envisage electron-proton ($ep$) center-of-mass (CM) energies $\sqrt{s}_{ep} \sim$ 20--100 GeV,
with possible extensions to higher energies, and aim to deliver luminosities 
$\sim 10^{34}$ cm$^{-2}$ s$^{-1}$ over the full energy range \cite{Accardi:2012qut,EIC-designs}. 
Acceleration of a wide variety of nuclear beams would be possible, ranging from 
the deuteron ($A = 2$) to heavy ions ($A \sim 200$).
In electron-nucleus ($eA$) scattering at the collider the CM energy per nucleon is 
lower compared to $ep$ 
by a factor $\sqrt{Z/A} \approx$ 0.7 (0.6) for light (heavy) nuclei, and the luminosity
per nucleon is approximately the same as in $ep$. Such a facility would
significantly expand the ``energy-luminosity frontier'' in electromagnetic scattering
(see figure~\ref{fig:elumi}), particularly for nuclei, and enable qualitative advances
in exploring short-range structure and Quantum Chromodynamics.
The EIC physics program includes studies of the nucleon's three-dimensional 
partonic structure (gluon spin, quark spin flavor decomposition, transverse momentum, 
spatial structure, correlations), the dynamics of color fields in nuclei (quarks/gluon 
densities in nuclei, nuclear shadowing, physics of high gluon densities), and the
conversion of color charge to hadrons (color transparency, parton propagation in
medium, hadronization) \cite{Accardi:2011mz,Boer:2011fh}.
%
%

The EIC would vastly expand the capabilities for studying heavy quark production in 
electromagnetic scattering as compared to present facilities. Measurements of 
open charm and beauty electro- and photoproduction were performed at the HERA $ep$ 
collider at $x_{\rm B} < 0.01$, using various methods of charm/beauty identification;
see \cite{Aaron:2010ib,Abramowicz:2013eja} and references therein. 
Open charm production was also observed 
at the COMPASS $\mu N$ fixed-target experiment \cite{Adolph:2012uz}. The EIC luminosity
is two orders-of-magnitude higher than that of HERA and would permit measurements of 
open charm/beauty production with much higher rates, extending the kinematic coverage
to the region of large $x_{\rm B}$ ($\sim 0.1$) and rare processes such as high-$p_T$ jets. 
Heavy quark production with electromagnetic probes could for the first time be measured 
on nuclear targets and used to study the gluonic structure of nuclei and the propagation 
of heavy quarks through cold nuclear matter with full control of the initial state. 
Next-generation detection capabilities at the EIC --- tracking, vertex detection,
and especially $\pi/K$ identification --- would open up new channels for charm/beauty 
reconstruction compared to HERA and further boost the rate of identified heavy quarks 
for physics purposes. The study of possible EIC applications to heavy quark production 
therefore deserves special attention. Here we consider heavy quark production as
a new method for measuring the gluon densities in nuclei.

\section{Heavy quarks as probe of nuclear gluons}
\begin{figure}
\begin{tabular}{ll}
\parbox[c]{0.3\textwidth}{\includegraphics[width=0.3\textwidth]{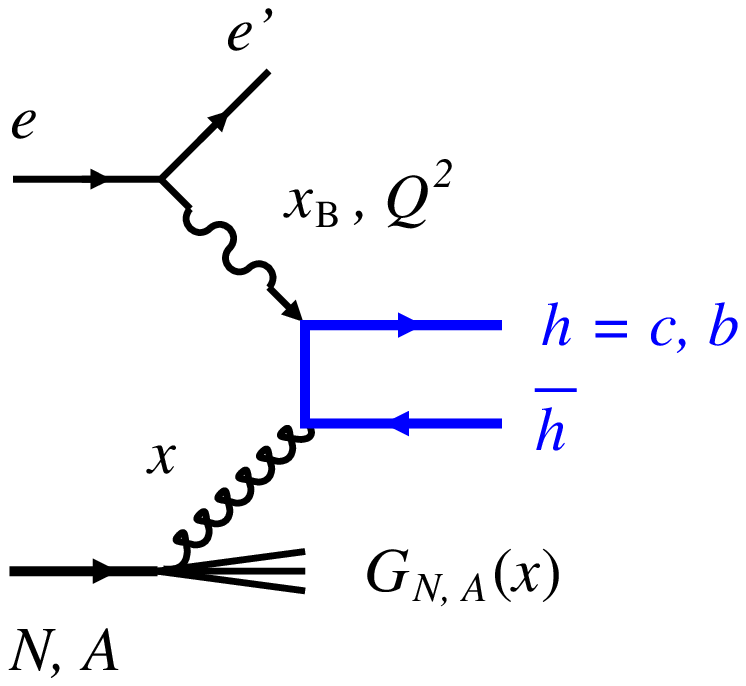}}
\hspace{0.05\textwidth}
&
\parbox[c]{0.55\textwidth}{
\includegraphics[width=0.55\textwidth]{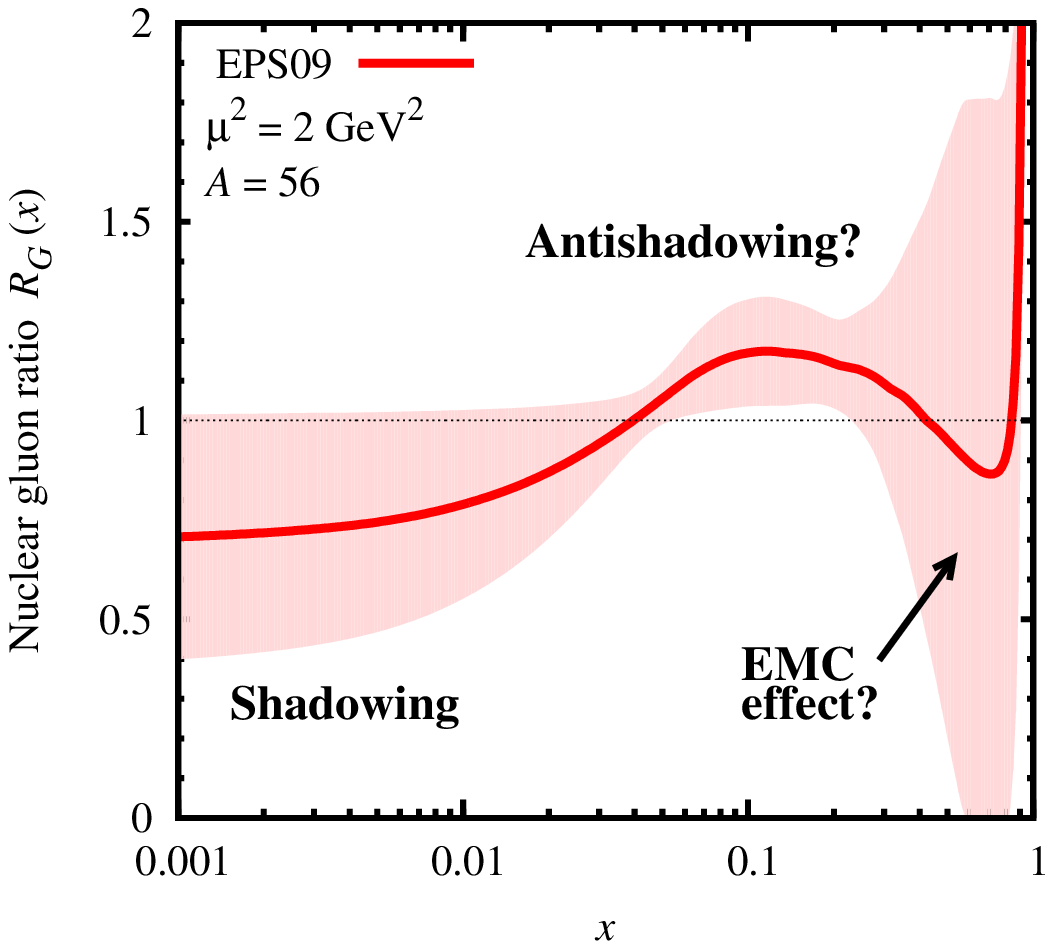}}
\\[-4ex]
{\small (a)} & {\small (b)}
\end{tabular}
\caption[]{(a) Heavy-quark production in DIS at LO (photon-gluon fusion).
(b) The nuclear gluon density ratio 
$R_G(x, \mu^2) = G_A(x, \mu^2) / [A G_N(x, \mu^2)]$ and its uncertainty
at a scale $\mu^2 = 2$ GeV$^2$, obtained from the 
EPS09 analysis of nuclear PDFs \cite{Eskola:2009uj}. 
\label{fig:nucl}
}
\end{figure}

Heavy quark production in DIS can serve as a direct probe of the gluon density in 
the target. At leading order in perturbative QCD the heavy quark pair is 
produced through photon-gluon fusion (see figure~\ref{fig:nucl}a) 
and samples the gluon density at momentum fractions $x > ax_{\rm B}$ ($x_{\rm B}$ 
is the Bjorken variable, $a = 1 + 4 m_h^2/Q^2$, and $m_h$ is the heavy quark mass), 
at an effective scale $\mu^2 \approx 4 m_h^2$; see \cite{Gluck:1993dpa} and
references therein. Higher-order QCD corrections
are known and theoretical uncertainties have been quantified \cite{Baines:2006uw}. 
The HERA results have shown good agreement with the QCD predictions \cite{Aaron:2010ib}. 
With the EIC heavy quark production could thus become a practical tool for measuring
the gluon densities in the nucleon and in nuclei.

Of particular interest is the gluon density in nuclei at large $x$. Measurements 
of inclusive DIS have shown that the valence quark densities in nuclei are 
suppressed at $x > 0.3$ (EMC effect) and inspired numerous theoretical studies of
QCD in nuclei \cite{Malace:2014uea}.
The nuclear modifications of gluons are largely unknown at present,
and basic questions remain to be answered (see figure~\ref{fig:nucl}b): Is the nuclear 
gluon density suppressed at $x > 0.3$ like the valence quarks (gluonic EMC effect)? 
Is the nuclear gluon density enhanced at $x \sim 0.1$ (gluon antishadowing)? 
The answers to these questions would offer insight into the change of the nucleon's 
gluonic structure due to nuclear binding and the QCD structure of 
nucleon-nucleon interactions. Information on the nuclear gluons at $x > 0.1$
can been obtained indirectly from the $Q^2$ dependence of inclusive nuclear 
DIS cross sections (DGLAP evolution), but the reach of the present 
fixed-target data is very limited. EIC would improve the situation by extending the
inclusive measurements over a larger $Q^2$ and $W$ range and separating 
longitudinal and transverse structure functions. A much more powerful method would be 
direct measurements of the nuclear gluon density at a fixed scale using heavy quark production.
%
%

Here we report about an R\&D project studying the feasibility of direct 
measurements of large-$x$ nuclear gluons using heavy quark production 
at EIC \cite{LD1601,Chudakov:2016otl}. 
The tasks include (a) estimating the charm production rates at EIC and
the angle-momentum distributions of the heavy mesons and their decay products;
(b) exploring new methods of charm reconstruction appropriate for large $x_{\rm B}$
using the EIC detector capabilities; (c) quantifying the impact on nuclear
gluons and the theoretical uncertainties. We note that a detailed 
design of the JLab EIC detector has yet to be completed, and that simulations of charm 
reconstruction methods at this stage are necessarily at the generic level.
While the studies of charm/beauty production reported here focus on the specific 
application to large-$x$ nuclear gluons, many of the results are more general 
and can be used for other physics studies with heavy quarks at EIC (heavy quark 
fragmentation functions, jets, propagation and hadronization in nuclei).

\section{Charm production in DIS at EIC}
\begin{figure}
\parbox[c]{0.64\textwidth}{
\includegraphics[width=0.64\textwidth]{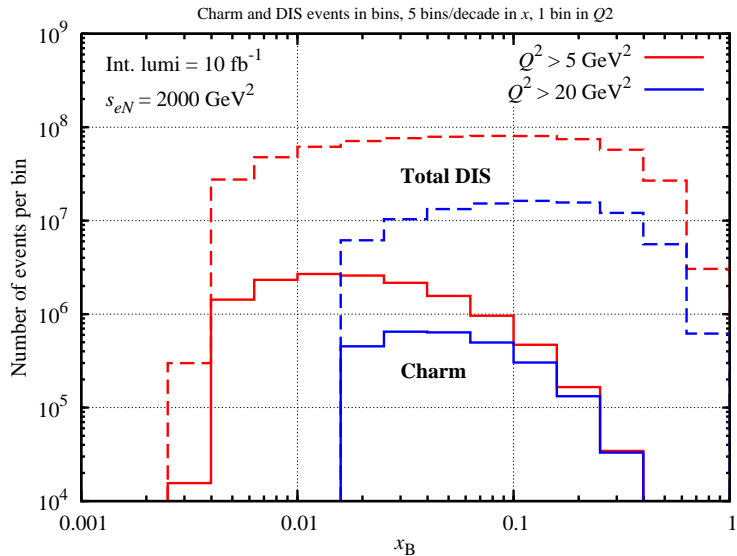}}
\hspace{0.02\textwidth} 
\parbox[c]{0.31\textwidth}{
\caption[]{Estimated number of DIS events (dashed lines) and charm events (solid lines) 
in DIS at EIC (CM energy $s_{eN} = 2000$ GeV$^2$, integrated nucleon luminosity 
10 fb$^{-1}$). The bins in $x_{\rm B}$ are 5 per decade as indicated on the plot. 
$Q^2$ is integrated from the lower value indicated (5 or 20 GeV$^2$) to the 
kinematic limit at the given $x_{\rm B}$.
\label{fig:rates}}}
\end{figure}

Charm production rates in DIS at EIC have been estimated using QCD
expressions and the HVQDIS code \cite{Harris:1997zq} (see figure~\ref{fig:rates}). 
The charm rates drop rapidly above $x_{\rm B} \sim 0.1$ 
due to the decrease of the gluon density. The fraction of DIS events with charm 
production changes from $\sim10\%$ at $x_{\rm B} \sim 0.01$ to $\sim1\%$ 
at $x_{\rm B} \sim 0.1$ (exact numbers depending on $Q^2$). The charm fraction
increases with $Q^2$ at fixed $x_{\rm B}$ as expected for a gluon-dominated process.
With an integrated luminosity of 10 fb$^{-1}$ charm production numbers of $\sim 10^{6}$ 
($\sim 10^{5}$) can be achieved in DIS at $x_{\rm B} \sim 0.01$ ($\sim 0.1$).
Higher charm rates could be achieved by lowering the $Q^2$ cutoff and/or including 
charm photoproduction. These numbers define the starting point for charm physics 
analysis. The challenge in gluon measurements at $x_{\rm B} \sim 0.1$ will be to identify 
charm events with an efficiency of $\sim$ few \%, in the presence of a DIS background
that is $\sim 100$ times larger.
%
%

The nucleon momenta in the medium-energy EIC proton/ion beams are 
$\sim$ few 10 GeV; e.g., collisions of 10 GeV electrons on 50 GeV nucleons 
at $s_{eN} =$ 2000 GeV$^2$. With this setup the charm quarks produced 
in DIS at $x_{\rm B} \sim 0.1$ typically emerge with large angles in the lab frame
(which approximately coincides with the virtual photon-gluon CM frame) and 
carry moderate momenta $\sim <$ 10 GeV. (The actual charm angle and momentum
distributions exhibit a complex dependence on $x_{\rm B}$ and $Q^2$ and
have to be determined by kinematic transformations.) The charm quark 
distributions are imparted on the produced charmed hadrons and their final decay 
hadrons ($\pi, K, p$). A major advantage of the medium-energy EIC design is that 
these hadrons emerge at large angles and moderate momenta, where good particle 
identification (PID), tracking, and vertex detection capabilities can be provided
by the central detector. In contrast, with a high-energy collider (HERA) the hadrons 
from large-$x_{\rm B}$ charm decays would appear at forward angles and with 
much larger momenta, rendering their detection more difficult.

\section{Toward charm reconstruction at EIC}
Charm events in DIS have to be identified through the charmed hadrons ($D$ mesons,
$\Lambda_c$ baryons) that are produced by charm quark fragmentation and subsequently 
decay into $\pi/K/p$, using either exclusive decay channels or inclusive modes (jets).

\begin{figure}[t]
\begin{center}
\includegraphics[width=0.95\textwidth]{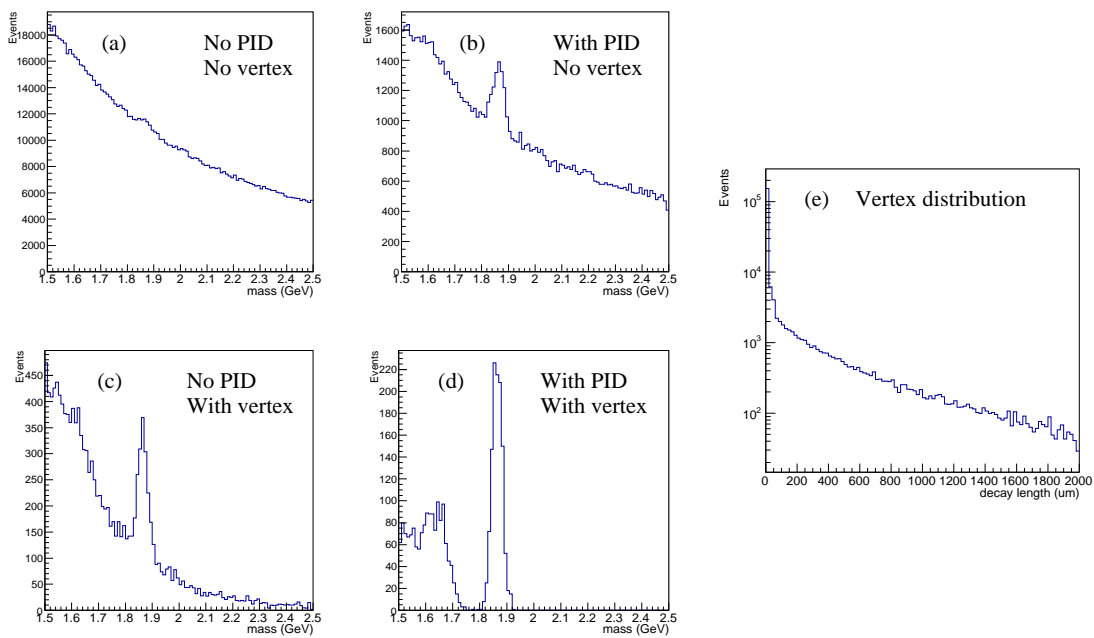}
\end{center}
\caption[]{Impact of PID and vertex cuts on $D^0$ meson reconstruction from 
the $K^- \pi^+$ exclusive decay. Plots (a)--(d) show the invariant mass spectrum 
of two charged tracks/mesons in a sample of charm events with $Q^2 > 10$ GeV$^2$
and $x_{\rm B} >$ 0.05 (PYTHIA 6 simulation, arbitrary normalization of event sample, 
background from non-charm DIS events not included). 
(a) No PID (charged tracks), 
no vertex cut; (b) with PID ($K^-, \pi^+$), no vertex cut; (c) no PID, with vertex cut; 
(d) with PID and vertex cut. Plot (e) shows the vertex distribution 
of the $D^0$ decay in the sample. The vertex cut was applied at 100 $\mu$m. 
\label{fig:pid_vertex}}
\end{figure}

A summary of significant exclusive decays of $D/\Lambda_c$ into charged $\pi/K$ is given
in table~\ref{tab:channels}. The theoretical reconstruction efficiency in this method
is determined by the product of the fragmentation ratio into the charmed hadron and
the branching ratio for the exclusive decay. HERA experiments made 
extensive use of the $D^{\ast +}$ channel, which exhibits a distinctive two-step 
decay  $D^{\ast +} \rightarrow D^0 \pi^+_{\rm slow}$, $D^0 \rightarrow K^- \pi^+$, and can 
be reconstructed without PID or vertex detection. However, this channel offers an 
overall reconstruction efficiency of only $\sim 1\%$, which is not sufficient 
for gluon measurements at $x_{\rm B} \sim 0.1$. The EIC detector will provide
PID capabilities for charged $\pi/K$ separation and permit use of other 
exclusive channels for charm reconstruction (see the example of $D^0$ reconstruction
through the $K^- \pi^+$ decay in figure~\ref{fig:pid_vertex}). Combining the charged 
exclusive decays in table~\ref{tab:channels} one could achieve a theoretical charm 
reconstruction efficiency of up to $\sim 10\%$, which would significantly expand the 
physics reach at large $x_{\rm B}$. 
Vertex detection can substantially improve the signal/background ratio in charm 
reconstruction through exclusive decays (see figure~\ref{fig:pid_vertex}),
but it reduces the overall reconstruction efficiency, because it rejects 
events with a short decay length. The optimization of charm reconstruction with exclusive
decays for the purpose of gluon measurements at $x_{\rm B} \sim 0.1$ at EIC is the object 
of on-going R\&D \cite{LD1601}.
%
%
\begin{table}
\parbox[c]{0.48\textwidth}{
\begin{tabular}{lrlr}
\br
$h_c$ & $f$ 
& Decay & BR \\
\hline
$D^0$         & 59\% & $K^-\pi^+$             & 3.9\% \\
              &      & $K^-\pi^+ \pi^+ \pi^-$ & 8.1\%   \\
$D^+$         & 23\% & $K^-\pi^+ \pi^+$       & 9.2\% \\
$D^{\ast +}$  & 23\% & $(K^-\pi^+)_{D0} \; \pi^+_{\rm slow}$ & 2.6\% \\
              &      & $(K^-\pi^+ \pi^+ \pi^-)_{D0} \; \pi^+_{\rm slow}$ & 5.5\%   \\
$D_s^+$       & 9\%  & $(K^+K^-)_\phi \; \pi^+$ & 2.3\% \\
$\Lambda_c^+$ & 8\%  & $p K^- \pi^+$ & 5.0\% \\
\br
\end{tabular}
}
\hspace{.03\textwidth}
\parbox[c]{0.47\textwidth}{
\caption{Channels for charm reconstruction through exclusive decays into charged hadronic 
final states. Columns: 1)~charmed hadron $h_c$; 2)~fragmentation fraction $f(c \rightarrow h_c)$;
3)~significant charged decay channels; 4)~branching ratio. The fragmentation fractions
are from ZEUS $\gamma p$ \cite{Abramowicz:2013eja}. They do not add up to 100\% because $D^0$ 
is also produced through $D^{\ast +}$ decays.
\label{tab:channels}}
}
\end{table}

Charm production in DIS can also be identified through inclusive modes, by selecting 
jet events with a displaced secondary vertex indicating a $D$ meson decay. A selection
method based on the decay length significance distribution was used in the last 
HERA experiments at $x_{\rm B} < 0.01$ \cite{Aaron:2010ib}. 
Such methods can in principle achieve much larger charm reconstruction efficiencies 
than exclusive channels, especially when combined with PID ($\sim$ 30\% was assumed in the EIC 
simulations in \cite{Accardi:2012qut}). Their feasibility for gluon measurements at 
$x_{\rm B} \sim 0.1$, where the fraction of charm events is only $\sim 1\%$ of DIS, 
needs to be explored further. Note that this assessment requires detailed assumption about 
the performance of the EIC vertex detector.

Another possible strategy for large-$x$ gluon measurements with charm is to focus 
on exceptional $c\bar c$ pairs with large transverse momenta $p_T \gg 1$ GeV.
While they are produced with a small cross section, such configurations represent 
a very distinctive final state that is practically free from hadronic background.
%
%

\section{Nuclear ratio measurements}
The application of charm production at EIC to the study of nuclear gluons 
at large $x$ also requires analysis of the uncertainties specific to nuclear
ratio measurements \cite{Chudakov:2016otl}. 
This includes (a) controlling the relative nuclear luminosity in
measurements with different ion beams through physics processes; (b) separating effects 
of nuclear final-state interactions on the observed meson spectrum from initial-state 
modifications of the nuclear gluon density using the different $A$-dependence of 
the two mechanisms; (c)~quantifying the impact of the charm production pseudodata 
on the nuclear gluon densities. Results will be reported in due course \cite{LD1601}.

\section{Summary}
A medium-energy EIC would offer excellent opportunities for measurements
of charm/beauty production in $ep/eA$ and $\gamma p/\gamma A$ scattering through 
a unique combination of energy, luminosity, and next-generation detection capabilities. 
The charm rates appear sufficient to constrain nuclear gluons at $x > 0.1$, if charm 
reconstruction could be performed with an overall efficiency of $\sim$ few \%.
Practical methods to achieve such efficiency are being investigated.
Heavy quark production at EIC could of course also be used for other physics purposes
(e.g.\ heavy quark fragmentation functions, jets, propagation and hadronization 
in nuclei), which may impose less stringent requirements. Simulating such 
measurements would elucidate other aspects of charm production/reconstruction with EIC
and provide further impulses for detector development.

This material is based upon work supported by the U.S.~Department of Energy, 
Laboratory Directed Research and Development funding, under contract DE-AC05-06OR23177.
The work of M.S.\ was supported by the U.~S.~DoE under award DE-FG02-93ER40771.
\end{document}